\documentstyle[12pt,amssym]{article}

\def\Ab{\bar{A}}
\def\diag{\mathop{\rm diag}\nolimits}
\def\sgn{\mathop{\rm sgn}\nolimits}
\def\R{\mbox{$\Bbb R$}}
\def\N{\mbox{$\Bbb N$}}
\def\sech{\mathop{\rm sech}\nolimits}
\def\arctanh{\mathop{\rm arctanh}\nolimits}

\title{
PT-SYMMETRIC NON-POLYNOMIAL OSCILLATORS AND HYPERBOLIC POTENTIAL WITH TWO
KNOWN REAL EIGENVALUES IN A SUSY FRAMEWORK}
\author{B. BAGCHI\\
{\small \sl Department of Applied Mathematics, University of Calcutta,} \\
{\small \sl 92 Acharya Prafulla Chandra Road, Calcutta 700 009, India}\\
{\small \sl E-mail: bbagchi@cucc.ernet.in}\\ [10pt] 
C. QUESNE\thanks{Directeur de recherches FNRS} \\
{\small \sl Physique Nucl\'eaire Th\'eorique et Physique Math\'ematique,}
\\ {\small \sl Universit\'e Libre de Bruxelles, Campus de la Plaine CP229,} \\ 
{\small \sl Boulevard~du Triomphe, B-1050 Brussels, Belgium} \\
{\small \sl E-mail: cquesne@ulb.ac.be}}
\date{ }
\begin{document}
\baselineskip=22pt plus 1pt minus 1pt
\maketitle

\begin{abstract}
Extending the supersymmetric method proposed by Tkachuk to the complex domain, we
obtain general expressions for superpotentials allowing generation of quasi-exactly
solvable PT-symmetric potentials with two known real eigenvalues (the ground state and
first-excited state energies). We construct examples, namely those of complexified
non-polynomial oscillators and of a complexified hyperbolic potential, to demonstrate how our
scheme works in practice. For the former we provide a connection with the sl(2) method,
illustrating the comparative advantages of the supersymmetric one. 
\end{abstract}

\vspace{1cm}
\noindent
Running head: SUSY and PT-symmetric potentials


\newpage
\section{Introduction}

Non-Hermitian Hamiltonians, in particular the PT-symmetric ones, are of great current
interest (see e.g.\ \cite{bender98a}--\cite{bagchi01c} and references quoted therein).
The main reason for this is that PT invariance, in a number of cases, leads to energy
eigenvalues that are real. In this regard, the early work of Bender and
Boettcher~\cite{bender98a} is noteworthy since it has sparked off some very interesting
developments subsequently.\par
%
%
Some years ago, Tkachuk~\cite{tkachuk98} proposed a supersymmetric (SUSY) method
for generating quasi-exactly solvable (QES) potentials with two known eigenstates, which
correspond to the wavefunctions of the ground state and the first excited state. A
distinctive feature of this method is that in contrast with other ones, it does not require
the knowledge of an initial QES potential for constructing a new one. Later on, the
procedure was extended to deal with QES potentials with two arbitrary
eigenstates~\cite{tkachuk01} or with three eigenstates~\cite{kuliy}. Quite recently,
Brihaye {\em et al.}~\cite{brihaye} established a connection between the Tkachuk
approach and the Turbiner one, based upon the finite-dimensional representations of
sl(2)~\cite{turbiner}.\par
%
%
In this letter, we pursue Tkachuk's ideas further by considering an extension of his results
to the case of PT-symmetric potentials. As a consequence, we arrive at a pair of
solutions, one of which is obtained by a straightforward and natural complexification of
Tkachuk's results while the other is new. We demonstrate the applicability of our scheme
by focussing on two specific potentials, both of which are PT-symmetric.\par
%
%
\section{Procedure}

\subsection{The Hermitian case}

Let us start with the Hermitian SUSY case, where the two known eigenstates are the
ground state and the first excited state. As is well known~\cite{cooper}, the SUSY
partner Hamiltonians are given by
\begin{equation}
  H^{(+)} = \Ab A = - \frac{d^2}{dx^2} + V^{(+)}(x), \qquad H^{(-)} = A \Ab = -
  \frac{d^2}{dx^2} + V^{(-)}(x),  
\end{equation}
for a vanishing factorization energy. Here $A$ and $\Ab$ are taken to be first-derivative
differential operators, namely
\begin{equation}
  A = \frac{d}{dx} + W(x), \qquad \Ab = - \frac{d}{dx} + W(x),  \label{eq:A}
\end{equation}
where $W(x)$ is the underlying superpotential and $V^{(\pm)}(x)$ are the usual SUSY
partner potentials
\begin{equation}
  V^{(\pm)}(x) = W^2(x) \mp W'(x).  \label{eq:V}
\end{equation}
\par
%
%
We assume SUSY to be unbroken with the ground state of the SUSY Hamiltonian $H_s
\equiv \diag(H^{(+)}, H^{(-)})$ to belong to $H^{(+)}$:
\begin{equation}
  H^{(+)} \psi^{(+)}_0(x) = 0, \qquad \psi^{(+)}_0(x) = C^{(+)}_0 \exp\left(- \int^x
  W(t) dt\right),  \label{eq:gs-wf} 
\end{equation}
$C^{(+)}_0$ being the normalization constant. Equation~(\ref{eq:gs-wf}) implies
\begin{equation}
  \sgn(W(\pm\infty)) = \pm 1.  \label{eq:sign}
\end{equation}
\par
%
%
In the above formulation of SUSY, the eigenvalues of $H^{(+)}$ and $H^{(-)}$ are
related as
\begin{equation}
  E^{(+)}_{n+1} = E^{(-)}_n, \qquad E^{(+)}_0 = 0,
\end{equation}
while the corresponding eigenfunctions are intertwined according to
\begin{equation}
  \psi^{(+)}_{n+1}(x) = \frac{1}{\sqrt{E^{(-)}_n}}\, \Ab \psi^{(-)}_n(x), \qquad 
  \psi^{(-)}_n(x) = \frac{1}{\sqrt{E^{(+)}_{n+1}}}\, A \psi^{(+)}_{n+1}(x). 
  \label{eq:psi-n}
\end{equation}
\par
%
%
{}Following Tkachuk~\cite{tkachuk98}, let us consider expressing $H^{(-)}$ in the form
\begin{equation}
  H^{(-)} = H^{(+)}_1 + \epsilon = A_1 \Ab_1 + \epsilon,
\end{equation}
where $\epsilon = E^{(+)}_1 = E^{(-)}_0$ corresponds to the energy of the first excited
state of $H^{(+)}$ or of the ground state of $H^{(-)}$ and the operators $A_1$ and
$\Ab_1$ are such that relations analogous to~(\ref{eq:A}) hold in terms of a new
superpotential $W_1(x)$. Thus we can write
\begin{equation}
  V^{(-)}(x) = V^{(+)}_1(x) + \epsilon  \label{eq:partner-V}
\end{equation}
with $V^{(+)}_1(x) = W_1^2(x) - W'_1(x)$ from~(\ref{eq:V}).\par
%
%
The ground state wave function of $H^{(+)}_1$ (or $H^{(-)}$) can be read off
from~(\ref{eq:gs-wf}) as
\begin{equation}
  \psi^{(-)}_0(x) = C^{(-)}_0 \exp\left(- \int^x W_1(t) dt\right), 
  \label{eq:partner-gs-wf}  
\end{equation}
where $C^{(-)}_0$ is the normalization constant and
\begin{equation}
  \sgn(W_1(\pm\infty)) = \pm 1.  \label{eq:partner-sign}
\end{equation}
On the other hand, the wave function of the first excited state of $H^{(+)}$ is obtained
in the form
\begin{equation}
  \psi^{(+)}_1(x) = C^{(+)}_1 \Ab \exp\left(- \int^x W_1(t) dt\right), \qquad
  C^{(+)}_1 = \frac{C^{(-)}_0}{\sqrt{\epsilon}},  \label{eq:exc-wf}  
\end{equation}
as guided by (\ref{eq:psi-n}) and (\ref{eq:partner-gs-wf}).\par
%
%
Using (\ref{eq:V}) and (\ref{eq:partner-V}), it is clear that the superpotentials $W(x)$
and $W_1(x)$ have to satisfy a constraint
\begin{equation}
  W^2(x) + W'(x) = W_1^2(x) - W'_1(x) + \epsilon.  \label{eq:constraint}
\end{equation}
The problem therefore amounts to finding a set of functions $W(x)$ and $W_1(x)$
satisfying (\ref{eq:constraint}) for some $\epsilon > 0$, along with the conditions
(\ref{eq:sign}) and (\ref{eq:partner-sign}). For this purpose, Tkachuk introduced the
following combinations
\begin{equation}
  W_{\pm}(x) = W_1(x) \pm W(x),  \label{eq:Wpm}
\end{equation}
which transform (\ref{eq:constraint}) into
\begin{equation}
  W'_+(x) = W_-(x) W_+(x) + \epsilon.  \label{eq:constraint-bis}
\end{equation}
An advantage with Eq.~(\ref{eq:constraint-bis}) is that it readily gives $W_-(x)$ in
terms of $W_+(x)$:
\begin{equation}
  W_-(x) = \frac{W'_+(x) - \epsilon}{W_+(x)}.  \label{eq:W-}
\end{equation}
Consequently the representations
\begin{equation}
  W(x) = \frac{1}{2} \left[W_+(x) - \frac{W'_+(x) - \epsilon}{W_+(x)}\right], \qquad
  W_1(x) = \frac{1}{2} \left[W_+(x) + \frac{W'_+(x) - \epsilon}{W_+(x)}\right] 
  \label{eq:W}
\end{equation}
satisfy Eq.~(\ref{eq:constraint}). In (\ref{eq:W}), $W_+(x)$ is some function for which
$W(x)$ and $W_1(x)$ fulfil conditions~(\ref{eq:sign}) and~(\ref{eq:partner-sign}). This
means
\begin{equation}
  \sgn(W_+(\pm\infty)) = \pm 1.
\end{equation}
\par
%
%
Restricting to continuous functions $W_+(x)$, the above condition reflects that
$W_+(x)$ must have at least one zero. Then from~(\ref{eq:W-}) and~(\ref{eq:W}),
$W_-(x)$, $W(x)$, and $W_1(x)$ may have poles. Tkachuk considered the case when
$W_+(x)$ has only one simple zero at $x = x_0$. In the neighbourhood of $x_0$, one gets
\begin{equation}
  \frac{W'_+(x) - \epsilon}{W_+(x)} \simeq \frac{W'_+(x_0) - \epsilon + (x-x_0)
  W''_+(x_0) + \cdots}{(x-x_0) W'_+(x_0) + \cdots},
\end{equation}
so that the superpotentials will be free of singularity if one chooses
\begin{equation}
  \epsilon = W'_+(x_0).  \label{eq:epsilon}
\end{equation}
One is then led to
\begin{equation}
  W(x) = \frac{1}{2} \left[W_+(x) - \frac{W'_+(x) - W'_+(x_0)}{W_+(x)}\right], \quad
  W_1(x) = \frac{1}{2} \left[W_+(x) + \frac{W'_+(x) - W'_+(x_0)}{W_+(x)}\right]. 
  \label{eq:W-bis}
\end{equation}
\par
%
%
To summarize, provided the continuous function $W_+(x)$ with a single pole at $x =
x_0$ is such that $W(x)$ and $W_1(x)$, given by~(\ref{eq:W-bis}), satisfy
conditions~(\ref{eq:sign}) and~(\ref{eq:partner-sign}), the Hamiltonian $H^{(+)}$ has
two known eigenvalues 0 and $\epsilon$, given by~(\ref{eq:epsilon}), with corresponding
eigenfunctions~(\ref{eq:gs-wf}) and
\begin{equation}
  \psi^{(+)}_1(x) = C^{(+)}_1 W_+(x) \exp\left(- \int^x W_1(t) dt\right).
  \label{eq:exc-wf-bis} 
\end{equation}
In deriving~(\ref{eq:exc-wf-bis}), use is made of~(\ref{eq:A}), (\ref{eq:exc-wf}),
and~(\ref{eq:Wpm}).\par
%
%
\subsection{The non-Hermitian case}

In the non-Hermitian case, we have to deal with complex potentials. Consider the
decomposition~\cite{andrianov}
\begin{equation}
  W(x) = f(x) + {\rm i} g(x), \qquad V^{(\pm)}(x) = V^{(\pm)}_R(x) + {\rm i} 
  V^{(\pm)}_I(x),  \label{eq:dec} 
\end{equation}
where $f$, $g$, $V^{(\pm)}_R$, $V^{(\pm)}_I \in \R$ and
\begin{equation}
  V^{(\pm)}_R = f^2 - g^2 \mp f', \qquad V^{(\pm)}_I = 2fg \mp g'.
\end{equation}
If $V^{(\pm)}(x)$ are PT-symmetric, then $f(x)$ and $g(x)$ are odd and even
functions, respectively.\par
%
%
It may be noted that Eqs.~(\ref{eq:gs-wf}) -- (\ref{eq:constraint}) remain true for some
real and positive $\epsilon$. Employing the separation
\begin{equation}
  W_1(x) = f_1(x) + {\rm i} g_1(x), \qquad V^{(+)}_1(x) = V^{(+)}_{1R}(x) +
  {\rm i} V^{(+)}_{1I}(x)  \label{eq:dec-bis} 
\end{equation}
with $f_1$, $g_1$, $V^{(+)}_{1R}$, $V^{(+)}_{1I} \in \R$, the behaviour of
$f_1(x)$ and $g_1(x)$ also turns out to be odd and even, respectively, should
$V^{(+)}(x)$ be PT-symmetric.\par
%
%
In the non-Hermitian case, the conditions~(\ref{eq:sign}) and~(\ref{eq:partner-sign}) are
to be replaced by
\begin{equation}
  \sgn(f(\pm\infty)) = \sgn(f_1(\pm\infty)) = \pm 1.  \label{eq:nH-sign}
\end{equation}
These conditions are actually compatible with the odd character of $f$ and $f_1$.\par
%
%
On introducing the first relations of~(\ref{eq:dec}) and~(\ref{eq:dec-bis})
into~(\ref{eq:constraint}) and splitting into real and imaginary parts, we get the system of
two equations
\begin{equation}
  f^2 - g^2 + f' = f_1^2 - g_1^2 - f_1' + \epsilon, \qquad 2fg + g' = 2f_1g_1 - g'_1.
  \label{eq:system}
\end{equation}
The problem now amounts to finding a pair of odd functions $f$, $f_1$ and a pair of
even functions $g$, $g_1$, satisfying Eq.~(\ref{eq:system}) for some $\epsilon > 0$, as
well as the conditions~(\ref{eq:nH-sign}).\par
%
%
To this end, we introduce the linear combinations
\begin{equation}
  f_{\pm}(x) = f_1(x) \pm f(x), \qquad g_{\pm}(x) = g_1(x) \pm g(x), 
\end{equation}
which replace the constraints~(\ref{eq:system}) by
\begin{equation}
  f'_+ = f_- f_+ - g_- g_+ + \epsilon, \qquad g'_+ = f_+ g_- + f_- g_+.
\end{equation}
Solving for $f_-$ and $g_-$ in terms of $f_+$ and $g_+$, we get
\begin{equation}
  f_- = \frac{(f'_+ - \epsilon) f_+ + g'_+ g_+}{f_+^2 + g_+^2}, \qquad
  g_- = \frac{- (f'_+ - \epsilon) g_+ + g'_+ f_+}{f_+^2 + g_+^2}.
\end{equation}
Hence the functions
\begin{eqnarray}
  f & = & \frac{1}{2} \left[f_+ - \frac{(f'_+ - \epsilon) f_+ + g'_+ g_+}{f_+^2 + g_+^2}
       \right], \qquad g = \frac{1}{2} \left[g_+ + \frac{(f'_+ - \epsilon) g_+ - g'_+
       f_+}{f_+^2 + g_+^2}\right], \nonumber \\
  f_1 & = & \frac{1}{2} \left[f_+ + \frac{(f'_+ - \epsilon) f_+ + g'_+ g_+}{f_+^2 +
       g_+^2}\right], \qquad g_1 = \frac{1}{2} \left[g_+ - \frac{(f'_+ - \epsilon) g_+ -
       g'_+ f_+}{f_+^2 + g_+^2}\right]  \label{eq:f-g}  
\end{eqnarray}
satisfy the coupled equations~(\ref{eq:system}). In~(\ref{eq:f-g}), $f_+$ must be such
that the conditions~(\ref{eq:nH-sign}) are fulfilled. These suggest
\begin{equation}
  \sgn(f_+(\pm\infty)) = \pm 1.  \label{eq:nH-sign-bis}
\end{equation}
\par
%
%
If we restrict ourselves to continuous functions $f_+(x)$ and $g_+(x)$, the
condition~(\ref{eq:nH-sign-bis}) shows that $f_+(x)$ must have at least one zero. For
simplicity's sake, we assume that $f_+(x)$ has only one simple zero at $x = x_0$. This
means that the parity operation is defined with respect to a mirror placed at $x = x_0$.
Thus in the neighbourhood of $x_0$, we get
\begin{eqnarray}
  \lefteqn{\frac{(f'_+ - \epsilon) f_+ + g'_+ g_+}{f_+^2 + g_+^2}} \\
  & \simeq & (x - x_0) \frac{[f'_+(x_0) - \epsilon] f'_+(x_0) + g_+(x_0)
       g''_+(x_0)}{[g_+(x_0)]^2} + \cdots \qquad \mbox{\rm if\ } g_+(x_0) \ne 0
       \nonumber \\
  & \simeq & \frac{1}{x - x_0} \frac{f'_+(x_0) - \epsilon}{f'_+(x_0)} + \cdots \qquad
       \mbox{\rm if\ } g_+(x_0) = 0, \\
  \lefteqn{\frac{(f'_+ - \epsilon) g_+ - g'_+ f_+}{f_+^2 + g_+^2}} \\
  & \simeq & \frac{f'_+(x_0) - \epsilon}{g_+(x_0)} + \cdots \qquad \mbox{\rm if\ }
      g_+(x_0) \ne 0 \nonumber \\
  & \simeq & - \frac{[f'_+(x_0) + \epsilon] g''_+(x_0)}{2 [f'_+(x_0)]^2} + \cdots
      \qquad \mbox{\rm if\ } g_+(x_0) = 0. 
\end{eqnarray}
Hence the superpotentials will be free of singularity if either $g_+(x_0) \ne 0$ and
$\epsilon$ arbitrary or $g_+(x_0) = 0$ and $\epsilon = f'_+(x_0)$ (or $\epsilon =
W'_+(x_0)$ since $g'_+(x_0) = 0$).\par
%
%
The superpotentials may therefore be written as
\begin{equation}
  W(x) = \frac{1}{2} \left[W_+(x) - \frac{W'_+(x) - \epsilon}{W_+(x)}\right], \qquad
  W_1(x) = \frac{1}{2} \left[W_+(x) + \frac{W'_+(x) - \epsilon}{W_+(x)}\right], 
  \label{eq:nH-W}
\end{equation}
if $W_+(x_0) = {\rm i} g_+(x_0) \ne 0$ and $\epsilon > 0$, or
\begin{equation}
  W(x) = \frac{1}{2} \left[W_+(x) - \frac{W'_+(x) - W'_+(x_0)}{W_+(x)}\right], \qquad
  W_1(x) = \frac{1}{2} \left[W_+(x) + \frac{W'_+(x) - W'_+(x_0)}{W_+(x)}\right], 
  \label{eq:nH-W-bis}
\end{equation}
if $W_+(x_0) = {\rm i} g_+(x_0) = 0$ and $\epsilon = W'_+(x_0)$.\par
%
%
To summarize, in the non-Hermitian PT-symmetric case, we get two types of solutions:
the one given by~(\ref{eq:nH-W-bis}) is obtained as a straightforward consequence of
the complexification of Tkachuk's result, while the other  given by~(\ref{eq:nH-W}) is
new.\par
%
%
\section{Applications}

\subsection{A family of complexified non-polynomial oscillators}

As the first application of our scheme we consider the case of a family of complexified
non-polynomial oscillators. This corresponds to the choice $f_+ = a x$, $g_+ = b
x^{2m}$ ($a>0$, $b \ne 0$, $m \in \N$), which leads to
\begin{equation}
  W_+ (x) = a x + {\rm i} b x^{2m}, \qquad m \in \N.
\end{equation}
Here $x_0 = 0$ and the first type of solutions, given in~(\ref{eq:nH-W}), applies to the
case $m=0$, while the second one, given in~(\ref{eq:nH-W-bis}), has to be used for $m
\in \N_0$.\par
%
%
The former case reduces to the well-studied exactly solvable PT-symmetric oscillator
potential~\cite{znojil99a, bagchi01b, bagchi01c}, thus partly accounting for the name of
the family of potentials. Indeed, to get it in the standard form we have to set $a=2$,
$b=-2c$, and $\epsilon = 4\alpha$. Then $V^{(+)}$ assumes the form
\begin{equation}
  V^{(+)}(x) = (x - {\rm i}c)^2 + 2 (\alpha - 1) + \frac{\alpha^2 - \frac{1}{4}}{(x - {\rm
  i}c)^2},  \label{eq:PT-osc}
\end{equation}
along with
\begin{equation}
  W(x) = x - {\rm i}c + \frac{\alpha - \frac{1}{2}}{x - {\rm i}c}, \qquad W_1(x) = x - {\rm
  i}c - \frac{\alpha - \frac{1}{2}}{x - {\rm i}c}. 
\end{equation}
These agree with the two independent forms of the complex superpotentials proposed by
us previously~\cite{bagchi01b} in connection with para-SUSY and second-derivative SUSY
of~(\ref{eq:PT-osc}). The potential~(\ref{eq:PT-osc}) can be looked upon as a
transformed three-dimensional radial oscillator for the complex shift $x \to x - {\rm i}c$,
$c>0$, and replacing the angular momentum parameter $l$ by $\alpha - \frac{1}{2}$.
The presence of a centrifugal-like core notwithstanding, the shift of the singularity off the
integration path makes (\ref{eq:PT-osc}) exactly solvable on the entire real line for any
$\alpha > 0$ like the harmonic oscillator to which (\ref{eq:PT-osc}) reduces for $\alpha
= \frac{1}{2}$ and $c=0$.\par
%
%
In contrast, the case $m \in \N_0$ is entirely new. For such $m$ values, the
superpotentials are given by
\begin{equation}
  W(x) = \frac{1}{2}\left[ax + {\rm i}bx^{2m} - \frac{2m{\rm i}bx^{2m-2}}{a + {\rm
  i}bx^{2m-1}}\right], \qquad W_1(x) = \frac{1}{2}\left[ax + {\rm i}bx^{2m} + 
\frac{2m{\rm i}bx^{2m-2}}{a + {\rm i}bx^{2m-1}}\right],  
\end{equation}
where we used $\epsilon = W'_+(0) = a$.\par
%
%
The corresponding potentials turn out to be
\begin{eqnarray}
  V^{(+)}(x) & = & \frac{1}{4} \Biggl[- b^2 x^{4m} + 2{\rm i}ab x^{2m+1} - 8m{\rm i}b
        x^{2m-1} + a^2 x^2 - 2a + \frac{4m(m-1){\rm i}b x^{2m-3}}{a + {\rm i}bx^{2m-1}}
        \nonumber \\
  && \mbox{} + \frac{4m(m-1){\rm i}ab x^{2m-3}}{(a + {\rm i}bx^{2m-1})^2}\Biggr], 
        \qquad m = 1, 2, 3, \ldots,  \label{eq:example1}
\end{eqnarray}
which are QES and are seen to be PT-symmetric as well. The ground and first-excited
state wave functions corresponding to the above QES potentials are easily determined
to be
\begin{eqnarray}
  \psi^{(+)}_0(x) & \propto & (a + {\rm i}bx^{2m-1})^{m/(2m-1)} \exp \left[- \frac{1}{4}
       ax^2 - \frac{{\rm i}b}{2(2m+1)} x^{2m+1}\right],  \label{eq:gs-wf1} \\
  \psi^{(+)}_1(x) & \propto & x (a + {\rm i}bx^{2m-1})^{(m-1)/(2m-1)} \exp \left[-
       \frac{1}{4} ax^2 - \frac{{\rm i}b}{2(2m+1)} x^{2m+1}\right].  \label{eq:fe-wf1}
\end{eqnarray}
\par
%
%
It is worth noting that the first member of the set~(\ref{eq:example1}) obtained for
$m=1$,
\begin{equation}
  V^{(+)}(x) = \frac{1}{4} \left(- b^2 x^4 + 2{\rm i}ab x^3 + a^2 x^2 - 8{\rm i}bx -
  2a\right),
\end{equation}
is a quartic potential differing from the known QES ones~\cite{bender98b, cannata}. All
the remaining members of the set, starting with that associated with $m=2$,
\begin{equation}
  V^{(+)}(x) = \frac{1}{4} \left[- b^2 x^8 + 2{\rm i}ab x^5 - 16{\rm i}b x^3 + a^2 x^2 - 2a
  + \frac{8{\rm i}bx}{a + {\rm i}bx^3} + \frac{8{\rm i}abx}{(a + {\rm i}bx^3)^2}\right], 
\end{equation}
are non-polynomial potentials. As for the PT-symmetric oscillator~(\ref{eq:PT-osc}),
the shift of the singularity off the integration path makes such potentials QES.\par
%
%
On introducing the new variable $z = x (a + {\rm i}bx^{2m-1})^{-1/(2m-1)}$, the
first-excited state wave function~(\ref{eq:fe-wf1}) can be rewritten in terms of the ground
state one~(\ref{eq:gs-wf1}) as $\psi^{(+)}_1(z) \propto z \psi^{(+)}_0(z)$. By setting in
general $\psi^{(+)}_n(z) = \psi^{(+)}_0(z) \phi^{(+)}_n(z)$, where $\phi^{(+)}_0(z)
\propto 1$ and $\phi^{(+)}_1(z) \propto z$, the Schr\"odinger equation for the
potentials~(\ref{eq:example1}) is transformed into the differential equation
\begin{equation}
  T \phi^{(+)}_n(z) \equiv \left[- a^{-2/(2m-1)} (1 - {\rm i}bz^{2m-1})^{4m/(2m-1)}
  \frac{d^2}{dz^2} + a z \frac{d}{dz}\right] \phi^{(+)}_n(z) = E^{(+)}_n \phi^{(+)}_n(z).
  \label{eq:T}
\end{equation}
\par
%
%
{}For $m=1$, the coefficient of the second-order differential operator in~(\ref{eq:T})
becomes a quartic polynomial in $z$, thus showing that $T$ can be expressed as a
quadratic combination of the three sl(2) generators
\begin{equation}
  J_+ = z^2 \frac{d}{dz} - Nz, \qquad J_0 = z \frac{d}{dz} - \frac{N}{2}, \qquad J_- =
  \frac{d}{dz},  \label{eq:sl(2)}
\end{equation}
corresponding to the two-dimensional irreducible representation (i.e.,
$N=1$ in~(\ref{eq:sl(2)}))~\cite{brihaye, turbiner}. The result reads
\begin{equation}
  T = a^{-2} \left(- b^4 J_+^2 - 4{\rm i}b^3 J_+ J_0 + 6b^2 J_+ J_- + 4{\rm i}b J_0 J_-
  - J_-^2 - 2{\rm i}b^3 J_+ + 6b^2 J_0 + 2{\rm i}b J_- + 3b^2\right).
\end{equation}
\par
%
%
{}For higher $m$ values, the differential operator $T$ contains a non-vanishing element
of the kernel~\cite{brihaye}. It is worth stressing that in such a case, the sl(2) method
becomes quite ineffective for constructing new QES potentials, whereas the SUSY one
is not subject to such restrictions.\par
%
%
\subsection{A complexified hyperbolic potential}

Our next example is that of a complexified hyperbolic potential induced by the
representations $f_+ = A \sinh \alpha x$, $g_+ = B$ ($A$, $\alpha > 0$, $B \ne 0$). We
then get
\begin{equation}
  W_+(x) = A \sinh \alpha x + {\rm i} B,
\end{equation}
which gives for $x_0=0$
\begin{eqnarray}
  W(x) & = & \frac{1}{2} \left[A\sinh\alpha x + {\rm i}B - \frac{A\alpha \cosh\alpha x -
        \epsilon}{A\sinh\alpha x + {\rm i}B}\right], \\
  W_1(x) & = & \frac{1}{2} \left[A\sinh\alpha x + {\rm i}B + \frac{A\alpha \cosh\alpha
        x - \epsilon}{A\sinh\alpha x + {\rm i}B}\right]. 
\end{eqnarray}
\par
%
%
The resulting expression for the complexified hyperbolic potential is
\begin{eqnarray}
  V^{(+)}(x) & = & \frac{1}{4} \Biggl[A^2 \sinh^2 \alpha x - 4A\alpha \cosh \alpha x +
        2\epsilon + \alpha^2 - B^2 + 2{\rm i}AB \sinh \alpha x \nonumber \\
  && \mbox{} + \frac{\epsilon^2 - \alpha^2(A^2-B^2)}{(A\sinh\alpha x + {\rm i}B)^2}
        \Biggr].  \label{eq:example2}
\end{eqnarray}
Clearly $V^{(+)}(x)$ is PT-symmetric. The two known eigenstates
of~(\ref{eq:example2}) correspond to the ground state and first excited state as
outlined earlier. These are 
\begin{eqnarray}
  \psi^{(+)}_0(x) & \propto & (A \cosh\alpha x - \nu)^{\frac{1}{4}\left(1 - \frac{\epsilon}
        {\alpha\nu}\right)} (A \cosh\alpha x + \nu)^{\frac{1}{4}\left(1 + \frac{\epsilon}
        {\alpha\nu}\right)} \nonumber \\
  && \mbox{} \times \exp\Biggl[- \frac{A}{2\alpha} \cosh\alpha x - \frac{1}{2} {\rm i}
        Bx - \frac{{\rm i}}{2} \arctan\left(\frac{A}{B} \sinh\alpha x\right) \nonumber \\
  && \mbox{} + \frac{{\rm i} \epsilon}{2\alpha\nu} \arctan\left(\frac{\nu}{B}
        \tanh\alpha x\right)\Biggr] \qquad \mbox{\rm if\ } 0 < B^2 < A^2, \nonumber \\
  & \propto & \cosh\frac{\alpha x}{2} \exp \left(- \frac{A}{2\alpha} \cosh\alpha x
        \right) \qquad \mbox{\rm if\ } B = 0, \nonumber \\
  & \propto & \sqrt{\cosh\alpha x} \exp\left[- \frac{A}{2\alpha} \cosh\alpha x -
       \frac{1}{2} {\rm i}\delta Ax + \frac{\epsilon}{2A\alpha} (\sech\alpha x + {\rm i}
       \delta \tanh\alpha x)\right] \nonumber \\
  && \qquad \mbox{\rm if\ } B^2 = A^2, \nonumber \\
  & \propto & (B^2 + A^2 \sinh^2 \alpha x)^{1/4} \exp \Biggl[- \frac{A}{2\alpha} \cosh
       \alpha x - \frac{\epsilon}{2\alpha\mu} \arctan\left(\frac{A\cosh\alpha x}{\mu}
       \right) \nonumber \\
  && \mbox{} - \frac{1}{2}{\rm i}Bx - \frac{{\rm i}}{2} \arctan \left(\frac{A}{B} \sinh
       \alpha x\right) + \frac{{\rm i}\epsilon}{2\alpha\mu} \arctanh\left(
       \frac{\mu}{B} \tanh\alpha x\right)\Biggr] \nonumber \\
  && \qquad \mbox{\rm if\ } B^2 > A^2,  \label{eq:example2-wf0} 
\end{eqnarray}
and
\begin{eqnarray}
  \psi^{(+)}_1(x) & \propto & (A \sinh\alpha x + {\rm i}B) (A \cosh\alpha x -
        \nu)^{- \frac{1}{4}\left(1 - \frac{\epsilon}{\alpha\nu}\right)} (A \cosh\alpha x +
        \nu)^{- \frac{1}{4}\left(1 + \frac{\epsilon}{\alpha\nu}\right)} \nonumber \\
  && \mbox{} \times \exp\Biggl[- \frac{A}{2\alpha} \cosh\alpha x - \frac{1}{2} {\rm i}
        Bx + \frac{{\rm i}}{2} \arctan\left(\frac{A}{B} \sinh\alpha x\right) \nonumber \\
  && \mbox{} + \frac{{\rm i} \epsilon}{2\alpha\nu} \arctan\left(\frac{\nu}{B}
        \tanh\alpha x\right)\Biggr] \qquad \mbox{\rm if\ } 0 < B^2 < A^2, \nonumber \\
  & \propto & \sinh\frac{\alpha x}{2} \exp \left(- \frac{A}{2\alpha} \cosh\alpha x
        \right) \qquad \mbox{\rm if\ } B = 0, \nonumber \\
  & \propto & (\sinh\alpha x + {\rm i}\delta) \sqrt{\sech\alpha x} \nonumber \\
  && \mbox{} \times \exp\Biggl[- \frac{A}{2\alpha} \cosh\alpha x - \frac{1}{2}
       {\rm  i}\delta Ax - \frac{\epsilon}{2A\alpha} (\sech\alpha x + {\rm i} \delta
       \tanh\alpha x)\Biggr] \qquad \mbox{\rm if\ } B^2 = A^2, \nonumber \\
  & \propto & (A\sinh\alpha x + {\rm i}B) (B^2 + A^2 \sinh^2 \alpha x)^{-1/4} \nonumber
       \\
  && \mbox{} \times \exp\Biggl[- \frac{A}{2\alpha} \cosh\alpha x +
       \frac{\epsilon}{2\alpha\mu} \arctan\left(\frac{A\cosh\alpha x}{\mu} \right)
       - \frac{1}{2}{\rm i}Bx \nonumber \\
  && \mbox{} + \frac{{\rm i}}{2} \arctan \left(\frac{A}{B} \sinh\alpha x\right) -
       \frac{{\rm i}\epsilon}{2\alpha\mu} \arctanh\left(\frac{\mu}{B}
       \tanh\alpha x\right)\Biggr] \qquad \mbox{\rm if\ } B^2 > A^2,  
       \label{eq:example2-wf1} 
\end{eqnarray}
where $\nu = \sqrt{A^2 - B^2}$, $\mu = \sqrt{B^2 - A^2}$, and $\delta = \sgn(B)$.\par
%
%
In (\ref{eq:example2-wf0}) and (\ref{eq:example2-wf1}), we have included the case $B=0$
for which the potential $V^{(+)}(x)$ of Eq.~(\ref{eq:example2}) reduces to one of the
potentials studied by Tkachuk~\cite{tkachuk98}, which itself is a special case of the Razavy
potential~\cite{razavy}.\par
%
%
\section{Conclusion}

To conclude, we have carried out in this paper a complexification of the SUSY method
proposed recently by Tkachuk. This allows us to generate QES PT-symmetric potentials with
two known real eigenvalues. We have also constructed suitable examples, namely those of a
family of complexified non-polynomial oscillators and of a complexified hyperbolic potential,
which serve to illustrate the viability of our scheme. For the former, we have also provided a
connection with the sl(2) method, which illustrates the comparative advantages of the SUSY
one.\par
%
%

\end{document}